\documentstyle[aps]{revtex}
\input psbox
\begin{document}
\draft
\title{\bf Anomalous Frequency-Dependent Conductivity near the Quantum Hall
Transition}
\author{ Giancarlo Jug\cite{GJ}$^a$ and Klaus Ziegler$^{a,b}$ }
\address{ $^a$Max-Planck-Institut f\"ur Physik Komplexer Systeme,
Au\ss enstelle Stuttgart, Postfach 800665, D-70506 Stuttgart (Germany) }
\address{ $^b$Institut f\"ur Physik, Universit\"at Augsburg, D-86135 Augsburg
(Germany)}
\date{\today}
\maketitle 
\begin{abstract}
The dynamical transport properties near the integer quantum Hall transition
are investigated at zero temperature by means of the Dirac fermion approach. 
These properties have been studied experimentally at low frequency $\omega$
and low temperature near the $\nu=1$ filling factor Hall transition, 
with the observation of an anusual broadening and an overall increase of the
longitudinal conductivity $Re \{ \sigma_{xx} \}$ as a function of $\omega$.
We find in our approach that, unlike for normal metals, the longitudinal 
conductivity increases as the frequency increases, whilst the width $\Delta B$
(or $\Delta\nu$) of the conductivity peak near the Hall transition increases.
These findings are 
in reasonable quantitative agreement with recent experiments by Engel et al. 
as well as with recent numerical work by Avishai and Luck.
\end{abstract}

\section{Introduction}

The two-dimensional (2D) electron gas placed in a strong perpendicular 
magnetic field exhibits, in the presence of a disordered one-particle 
potential, the exact quantization of the Hall conductivity $\sigma_{xy}$ 
accompanied by the vanishing of the longitudinal conductivity $\sigma_{xx}$. 
Whilst this integer quantum Hall effect (IQHE) is now fairly well understood 
\cite{revs}, the transition region between consecutive Hall plateaus has 
recently attracted a good deal of attention by the condensed-matter physics 
community. Extensive experimental \cite{wei,yamane,mceuen,sha1,rokhinson}, 
theoretical \cite{hikami,ludwigetal,zie2,zie6} as well as numerical 
\cite{chalk,huck} work has been carried out to characterize the transport 
properties in the transition region. This region arises due to the existence
of delocalized electronic states responsible for the jump in the Hall 
conductivity between two consecutive Hall plateaus. Within this region 
$\sigma_{xx}$ takes up a narrow peak-like form.

The detailed study of the transition region between quantum Hall plateaus
is important in testing and enhancing our understanding of the IQHE and of
the underlying physics of localization in the presence of a magnetic field.
In this paper we contribute to the advance of the theoretical description
by showing how the Dirac fermion approach \cite{ludwigetal,hal,hats,osh}, 
a successful theoretical treatment in the description of the IQH 
transition, affords a detailed explanation of some puzzling recent 
experimental measurements by Engel et al. \cite{engel} and Balaban et al.
\cite{balaban} of the 
frequency-dependent (AC) conductivity in the IQH system at low temperatures. 

Hitherto, much work has concentrated on the characterization of the static 
transport properties in the neighborhood of the transition region, where the
approach to the localization-delocalization transition is dominated by a 
diverging localization length $\xi{\sim}|E-E_c|^{-\nu}$. More recently, some
attention has been devoted to frequency-dependent properties, where the 
approach to the critical point $E_c$ is characterized also by diverging time
scales. The frequency-dependent AC conductivity is an interesting and probably
characteristic quantity of a 2D conducting QH system where $\sigma_{xx}$ is 
non-zero (metallic regime). In contrast to normal metals, where a Drude-like 
behavior is observed, the conductivity of a 2D QH system has been reported 
to display an {\it increasing} behavior with frequency \cite{engel}. Thus it 
is indicative of a new class of systems, different from normal metals (for 
which the conductivity {\it decreases} with frequency, or temperature). This 
can be seen from the peak-shape of $\sigma_{xx}(B,\omega)$ as a function of
the magnetic field $B$ in the recent IQHE measurements of Ref. \cite{engel}: 
the width $\Delta B$ of the conductivity peak broadens as the frequency (or 
temperature) is increased. 
Numerical simulations, based on the lowest Landau level projection, initially 
gave indications only for a decreasing (yet non-Drude like) 
$\sigma_{xx}(\omega)$ \cite{brenig}. 
However, a more recent numerical investigation by Avishai and 
Luck \cite{avishai} has provided strong evidence for a broadening of the 
longitudinal conductivity peak with frequency. 

This is a convenient point where to make the case for the theoretical 
approach -- based on Dirac fermions -- that is to be used in this paper and 
also to put it into the right perspective. Most -- although by all means not
all \cite{revs,ferrari} -- of the theoretical schemes that have been so 
far developed to explain the IQHE on a microscopic basis, rely on the
concept of localization of the single-electron states in the presence of 
any finite amount of disorder. Generally speaking, it is sufficient to have
a finite but narrow band of extended states near the center of each Landau
level, and total localization everywhere else in the density of states (DOS),
to explain most features of
transport in the IQHE. A detailed theory, stemming from the field-theoretic
and scaling approach to Anderson localization {\it in the absence} of a
magnetic field, has been developed by Pruisken et al. \cite{llp,pruisk} and 
rewarded with some experimental evidence for its correctness \cite{wei1}. 
This approach makes use of the concept of composition of the $\sigma_{xx}$
and $\sigma_{xy}$ conductivities and leads to a scaling theory where these
conductivities are both universal (although the actual value of $\sigma_{xx}$
is yet undetermined exactly at the transition) and the degenerate
extended states at each transition are concentrated into single-energy 
points, say in the DOS. Outside these points, corresponding
to the free-electron Landau energy levels, there is total
localization, although in the original paper by Levine et al. \cite{llp}
the possibility of a {\it band} of extended states near the plateau-to-plateau
transition could not be ruled out.
This approach certainly represents a useful picture for most
numerical as well as experimental studies of the IQH transition, leading
to the concept of quantum critical point that may prove useful also to 
other areas of research in condensed-matter theory. 
The approach of Pruisken et al. \cite{llp,pruisk} certainly remains an
appealing global picture
reproducing the pattern of the IQH transport experiment. However, our point of
view here is that the single-energy extended-states picture cannot be the
full story in a detailed microscopic understanding of the IQH transition.
Infinitely-narrow single-particle levels are a paradigm for bound states
(point spectrum),
not for extended ones (continuous spectrum). This is in line with Heisenberg's
indetermination principle; for if a quantum particle can be made to sample a
band of extended states with energy width $W$ (via e.g. an infinitesimal 
disturbance) it will acquire a momentum uncertainty

\begin{equation}
\Delta p =\sqrt{2\mu W},
\end{equation}

\noindent
$\mu$ being some effective inertial mass (not just in the sense of periodic
potential's band structure). If we take 
the disordered potential to be
characterized by a spatial-correlation characteristic length $\ell$, then
$\Delta p\ell \sim \hbar$ yields

\begin{equation}
\mu \sim \hbar^2/(2W\ell^2). 
\end{equation}

\noindent
So, an infinitely narrow band would lead to 
an infinite effective mass or zero mobility and no conduction. This argument
-- though by no means a proof -- works for the atomic as well as for the 
perfect 
periodic lattice limits. It is also important to understand conduction in 
doped semiconductors, where infinitely narrow impurity levels represent 
weakly-bound localized states giving way to an impurity conduction band only 
in the limit of heavily doped samples. The outstanding example we know of a
single-energy {\it localized} degenerate state is that of a Landau level in
the impurity-free electron gas. There, the enormous degeneracy of the Landau
levels is a consequence of the independence of energy on the orbital momentum
$L_z$. In our view, as soon as impurity collisions take place there will
be new states, some of them extended and ready to accept the scattered
electrons. The finite band width of such states ensures the mobility of the
electrons being
scattered and therefore conduction. The finite band-width picture for the
extended states would thus seem to be in agreement with basic quantum physics
principles.
We should stress, however,  that there exist 1D models of electrons in 
random potentials
\cite{hikami,ludwigetal,lifshits,dunlap,bovier,flores} which do demonstrate 
the possibility of
single energy extended states. Nevertheless, these models have a singular DOS
(vanishing or divergent at this energy), a situation which is not covered 
by the above qualitative argument. In this paper we aim at describing physical
2D systems with a regular DOS and no extreme behavior.

The Dirac fermion model with an inhomogeneous mass, as used in this paper,
is a plausible representation for electrons undergoing the IQH transition 
in the presence of a random potential,  as discussed in Ref.
\cite{ludwigetal}. 
As explained also in the next Section,
initial work with this model relied on perturbative calculations which led
to unphysical results for the density of states \cite{ludwigetal}. However, 
it was shown by one of the present Authors that a non-perturbative approach 
would cure this problem yielding all the desired features with the sole
random mass ingredient. Moreover, the non-perturbative calculation leads
to a narrow, but finite band width for the extended states 
(\cite{zie6,zie1}, see also \cite{leewa}). As we have explained, we find
this feature of the model rather attractive, together with the fact that the 
approach lends itself to a number of predictive analytical calculations not
possible
otherwise with the single-energy picture. The Dirac fermion formulation 
was recently also strengthened by a mapping of the network model \cite{chalk}
for the IQH transition onto a Dirac effective Hamiltonian with both
inhomogeneous mass and scalar as well as vector potentials \cite{ho}.
In our view, the inhomogeneous mass is sufficient to give the full picture,
and at the same time the simplest.

The purpose of this paper is to investigate the low-frequency behavior of the 
conductivity peak on the basis of a theory accounting directly for the IQH 
delocalized states, using this effective and appealing approach based on Dirac 
fermions. 
This approach has so far been used to account for the static transport 
properties of the IQH transition, and the method can be extended to include 
thermal fluctuations \cite{zie7}. Frequency-dependent behavior is, however, 
very similar to temperature behavior \cite{engel}, since dynamics enters 
in the formalism through the Matsubara frequencies, which are themselves 
proportional to temperature. We will therefore work only with the
frequency-dependent Dirac fermion approach. A complementary approach to
the frequency-dependent conductivity was worked out by Polyakov and
Shklovskii using the hopping mechanism of localized states
\cite{shklovskii}. They derived power laws for the broadening of the
conductivity peaks due to frequency, current and temperature. In contrast
to our approach, which works very close to the conductivity peak at low
frequency, they study the broadening in the regime of higher frequency
where the electronic states become localized if the frequency goes to
zero. The success of this hopping approach calls for an extension of these
ideas to the regime where the states are delocalized. This is precisely what
we shall do in the following using the method of Dirac fermions which seems
to afford a good deal of predictive power.

The paper is organized as follows. In Section II we briefly recall the basic
features of the Dirac fermion approach to the IQH transition. This approach
is implemented in Section III in order to directly evaluate the AC 
longitudinal 
conductivity $\sigma_{xx}(\omega)$ from the Dirac fermion propagator in
which a weakly-disordered one-particle potential is accounted for. The main
features of the AC conductivity are described in the light of recent numerical
as well as experimental investigations of dynamical scaling in the IQH 
transition region. In Section IV the description is specialized for the
frequency dependence of the conductivity peak width, which is found to be 
in agreement with some recent measurements by Shahar et al. \cite{sha2}
carried out for the (related) temperature dependence of the longitudinal
resistivity width and by Balaban et al.\cite{balaban} for the frequency
dependence.
For non-vanishing frequencies the results of
our calculation also agree with the experimental data of Engel et al.
\cite{engel}
and the numerical data of Avishai and Luck \cite{avishai}. Section V 
contains our conclusions and outlook.    


\section{The Model and Derivation of the Conductivity}

The main features of the IQH transition are captured quite effectively by a 
tight-binding model in which, although the real system has no lattice, the 
electrons hop over a scale given by the magnetic length \cite{ludwigetal}. 
This is closely 
related to the Chalker-Coddington network model \cite{chalk} in which electrons
hop from region to region with random tunneling and random magnetic flux.
Both models \cite{ludwigetal,ho} lead in the large-scale approximation to an
effective Hamiltonian describing the dynamics of Dirac fermions (with a random 
mass or coupled to a random vector potential) 

\begin{equation}
H_D=(i\nabla_1+A_1)\sigma_1+(i\nabla_2+A_2)\sigma_2+M\sigma_3,
\label{hamiltonian}
\end{equation}

\noindent
where the energy is measured in units of the hopping parameter $t$ of the
original lattice model. 
$\nabla_j$ is the lattice differential operator in the $j$-direction and
$\{\sigma_j\}$ are the Pauli matrices.
This Hamiltonian, with a random mass term $M$ and zero random vector potential
$A_j$, is a reasonable starting point 
for the description of the IQH transition between plateaus at filling $\nu=1$.
One important issue consists in what type of randomness is realistic in the
approach with Dirac fermions. Ludwig et al. argued that the case of a random 
Dirac mass is insufficient to describe the generic IQH transition since it has
a vanishing DOS at the transition point. Yet the random Dirac mass is
reasonable on an intuitive basis 
because it is the most relevant random contribution to the Dirac Hamiltonian
in terms of symmetry breaking \cite{remark} (see discussion in Ref. 
\cite{ludwigetal}).
However, it was shown by several other Authors 
and approaches that the DOS becomes non-zero at the transition when treated 
on a non-perturbative basis \cite{zie6,zie1,leewa}. Therefore the random mass 
case, in contrast to that of a pure random vector potential, ought to 
represent a generic model 
for the IQH transition and this is the point of view adopted in this article.
We choose a random mass $M=m+\delta M$ with mean $m$ and a Gaussian 
distribution with
$\langle \delta M_r\delta M_{r'} \rangle=g\delta_{rr'}$, $g$ being a measure
of the strength of disorder.\\


The frequency-dependent conductivity at $T=0$ reads \cite{wegner}

\begin{equation}
\sigma_{xx}(\omega)={\omega\over2}\int_{-\omega}^\omega{\hat\sigma}_{xx}
(\omega,E)dE
\label{cond0}
\end{equation}

\noindent
with

\begin{equation}
{\hat\sigma}_{xx}(\omega,E)
=-{e^2\over h}\lim_{\eta\to0}\sum_r r^2\langle G(0,r;E+\omega+i\eta)
G(r,0;E-\omega-i\eta)\rangle
\equiv{e^2\over h}\lim_{\eta\to0}\nabla_k^2
{\tilde C}(k,\eta-i\omega,E)|_{k=0},
\label{cond1}
\end{equation}

\noindent      
where $G$ is the one-particle Green's function of $H_D$.
In the following we will use the standard approximation of small $\omega$ for
$\sigma_{xx}(\omega)$ \cite{hikami}

\begin{equation}
\sigma_{xx}(\omega)\approx\omega^2{\hat\sigma}_{xx}(\omega,0).
\label{cond2}
\end{equation}

\noindent
This approximation will be shown to be equivalent to the Einstein relation, as
can be deduced from the expression for ${\tilde C}(k,\eta-i\omega,E)$ which 
will be derived below.

According to the Dirac fermion approach of Ref. \cite{zie6}, the two-particle
Green's function describes a diffusive behavior between the Hall plateaus

\begin{equation}
{\tilde C}(k,\eta-i\omega,0)=
{\pi\over2}{\rho\over \eta-i\omega+Dk^2},
\label{cricor}
\end{equation}

\noindent
where $\rho$ is the average DOS and $D$ the diffusion coefficient (notice that
we use a notation for $D$ different from that of Ref. \cite{zie6})

\begin{equation}
D=2g\eta'\alpha\Big[
1+\alpha({(m'+i\eta')^2\over1/g-2\alpha (m'+i\eta')^2}
+{(m'-i\eta')^2\over1/g-2\alpha(m'-i\eta')^2})\Big]
\label{diff0}
\end{equation}

\noindent
with (for an infinite cutoff)

\begin{equation}
\alpha=\int({m'}^2+{\eta'}^2+k^2)^{-2}d^2k/4\pi^2= 
{1\over4\pi({m'}^2+{\eta'}^2)}.
\end{equation}

\noindent
The parameters $m'$ and $\eta'\equiv \pi g\rho$ have been evaluated within
a saddle point approximation \cite{zie6}. They obey the following equations
(taking the limit $\eta\to0$)

\begin{equation}
\eta'+i\omega =\eta' g I
\label{spea}
\end{equation}

\noindent
and 

\begin{equation}
m'=m/(1+gI)
\label{speb}
\end{equation}

\noindent
with the integral $I$ given by

\begin{equation}
I\sim {1 \over \pi}\int_0^\lambda({\eta'}^2+{m'}^2+k^2)^{-1}kdk
= {1\over 2\pi}\ln\lbrack 1+\lambda^2/({\eta'}^2 +{m'}^2)\rbrack.
\end{equation}
Here we have cut-off the $k$-integration to $|k|\le\lambda$. This is
necessary because the integral $I$ would not otherwise exist. The cut-off
corresponds to a minimal length scale in the real system, i.e. the lattice
constant in our model, which is usually
the mean free path of the particles. It reflects the fact that
(quasi)particles cannot be considered as independent on arbitrary short
scales. 

\section{Evaluation of the Frequency-Dependent Conductivity}

Assuming weak disorder ($g\ll 1$), we get from Eq. (\ref{diff0})

\begin{equation}
D\approx 2g\eta'\alpha[1+2g\alpha({m'}^2-{\eta'}^2)]
={g\eta'\over2\pi({m'}^2+{\eta'}^2)}
[1+{g\over2\pi}{{m'}^2-{\eta'}^2\over{m'}^2+{\eta'}^2}],
\label{diff1}
\end{equation}

\noindent
which for the special case $m=0$ becomes 
\begin{equation}
D={g\over2\pi\eta'}(1-g/2\pi)={1\over2\pi^2\rho}(1-g/2\pi).
\label{D'0}
\end{equation}

\noindent
Then the conductivity reads, according to Eqs. (\ref{cond1}), (\ref{cond2})
and (\ref{cricor})

\begin{equation}
\sigma_{xx}(\omega)\approx{e^2\over \hbar} D\rho.
\label{cond1a}
\end{equation}

\noindent
We notice that this is the Einstein relation. Going back to the general
case,
Eqs. (\ref{diff1}) and (\ref{cond1a}) imply for the conductivity the
following expression

\begin{equation}
\sigma_{xx}(\omega)\approx
{e^2\over h\pi}{1\over 1+{m'}^2/{\eta'}^2}
[1+{g\over2\pi}{{m'}^2/{\eta'}^2-1\over{m'}^2/{\eta'}^2+1}].
\label{accond}
\end{equation}

\noindent
This represents a simple scaling form of the type

\begin{equation}
\sigma_{xx}(m,\omega)={e^2\over h}G(m'/\eta'),
\end{equation}

\noindent
since only the combination $m'/\eta'$ enters the expression.\\

Now we have to evaluate $m'$ and $\eta'$ from Eqs. (\ref{spea}) and
(\ref{speb}). For $\omega,\eta'\ne 0$ Eq. (\ref{spea}) can also be written as

\begin{equation}
e^{2\pi/g}e^{2\pi i\omega/\eta'g}=1+{\lambda^2\over {\eta'}^2+{m'}^2},
\label{spea1}
\end{equation}

\noindent
and since we are interested in the small frequency regime $\omega\approx0$
we find it useful to work out a closed approximate analytic solution of this
equation. For $\omega\approx0$, we have $gI\approx1$, from Eq. (\ref{spea}).
Then Eq. (\ref{speb}) implies $m'\approx m/2$ and the exponential term in
Eq. (\ref{spea1}) can be  expanded to give, in leading order

\begin{equation}
{\eta'}^2\approx\lambda^2e^{-2\pi/g}(1-{2\pi\over g}i{\omega\over\eta'})
-m^2/4={m_c^2-m^2\over4}-{2\pi i m_c^2\over4g}{\omega\over\eta'}
\label{cubic}
\end{equation}

\noindent
with $m_c=2\lambda\exp(-\pi/g)$. This is a cubic equation in $\eta'$ from 
which we take the solution

\begin{equation}
\eta'=\cases{
y^{1/3}+{a\over 3}y^{-1/3} & $a<-3(b^2/4)^{1/3}$ \cr
-(1/2)[y^{1/3}+{a\over 3}y^{-1/3}
+i\sqrt{3}(y^{1/3}-{a\over 3}y^{-1/3})] & $a\ge-3(b^2/4)^{1/3}$ \cr
}
\end{equation}

\noindent
with

\begin{eqnarray}
y & = & -i(b/2+\sqrt{b^2/4+a^3/27})
\nonumber\\
a & = & {m_c^2-m^2\over4},\qquad b= \frac{2\pi m_c^2}{4g}\omega .
\end{eqnarray}

\noindent
This solution reproduces the correct result in the limit $\omega\to0$,
namely $\eta'\to\sqrt{a}$ for $a\ge0$ and $\eta'\to0$ for $a<0$ \cite{zie7}.
Using the approximate values of $m'$ and $\eta'$ from above we can study

\begin{equation}
\sigma_{xx}(\omega)\approx{e^2\over h\pi}{1\over 1+m^2/4{\eta'}^2}.
\label{cond3}
\end{equation}

\noindent
For $b\propto\omega$ large compared to $(m_c^2-m^2)/4$ we have $\eta'\propto
\omega^{1/3}$ from Eq. (\ref{cubic}). Consequently, the scaling behavior
of the conductivity is given in this regime by

\begin{equation}
\sigma_{xx}(m,\omega)={e^2\over h} G(m\omega^{-1/3}).
\label{scaling1}
\end{equation}

\noindent
This, of course, does not hold for all values of $\omega$ because $\eta'$ is
not a power law for very small $\omega$. The general behavior of the real and 
imaginary parts of $\sigma_{xx}(\omega)$ is shown in Fig. 1, for the 
illustrative values $m_c=0.01$ and $m=0.009$.
\\

\section{Broadening of the $\sigma_{xx}$ peak}

Much of the experimental and numerical work on this problem has concentrated 
on the conjectured universal scaling behavior of the peak width of 
$\sigma_{xx}(B,\omega)$. 
It is clear from our calculation in the previous Section that the peak width of
our model does not vanish with vanishing frequency. There are, at this 
point, two possible attitudes for this fact: 
either our model does not capture the physics of the real systems, or the
non-vanishing width is too small to be resolved in the experiments \cite{wei1}
or in the computer calculations \cite{chalk,huck}. 
There are alternative models, for instance the
2D Dirac fermions with random vector potential in place of the random mass,
which do have a vanishing peak width \cite{ludwigetal}. However, these models
have an unphysical
behavior due to a singular DOS and a peak height different from experimental
observations \cite{rokhinson}. Since within a non-perturbative calculation
the peak height and the smooth behavior of the DOS in the case of a random mass
are in good agreement with experiments, it is more likely that
the second point of view applies to our model. This we have advocated for in
our presentation of the theoretical framework used in this paper, 
as given in the Introduction. Moreover, very recent
experiments indicate that the peak width does indeed not vanish
in the zero temperature and zero frequency limit \cite{sha2,balaban}.
The exponential dependence of the width on the disorder parameter may
explain why it is difficult to measure the narrow peak width, in particular in
samples with weak disorder.

The broadening of the peak width can be seen in Fig.2, where $Re(\sigma_{xx})$
is plotted as a function of the average Dirac mass $m$ and the frequency
$\omega$ for, as an illustration, disorder strength $g=0.6$ corresponding to
the value $m_c=0.01$. For
$m\approx m_c$ the conductivity $Re(\sigma_{xx})$ varies roughly like
$\omega^{2/3}$ as one would expect from Eqs. (\ref{cond3}) and 
(\ref{scaling1}).
However, away from $m_c$ the broadening does not, strictly speaking, describe
a power law.

\section{Discussion and Conclusions}

The broadening of the conductivity peak has also been studied numerically by
Avishai and Luck \cite{avishai}. They studied the scaling of the real part of 
the conductivity $\sigma_{xx}(E,\omega)$, finding that there is indeed
a scaling law of the type

\begin{equation}
Re \sigma_{xx}(E,\omega)\approx{e^2\over h}G(|E-E_c|\omega^{-\kappa})
\label{scaling}
\end{equation}

\noindent
for the real part of the conductivity. These Authors found values for 
$\kappa$ between 0.31 and 0.43. We can use our result from Eq. (\ref{accond}), 
where  $E-E_c$ is replaced by the parameter $m$ driving the IQH transition. 
However, 
in contrast to Eq. (\ref{scaling}), we do not obtain a simple scaling form for
arbitrarily small frequency (see Fig.2).
Still, for weak disorder the agreement with Ref. \cite{avishai} is
qualitatively
reasonable, at least for small values of $m-m_c$ (the numerical values quoted
are for the critical point).

In general, if we insist on a scaling law, the exponent $\kappa$ is related to
the dynamical exponent
$z$ and to the localization length exponent $\nu$ by $\kappa=1/z\nu$.
From $\kappa=1/3$ we could evaluate $\nu$ because
the usual argument for the dynamical exponent of non-interacting particles
is $z=2$ \cite{avishai}. The latter arises from the assumption that the
diffusion coefficient and the DOS are constant near the critical point in
the two-particle Green's function of Eq. (\ref{cricor}).  However, if the
DOS itself behaves like a power law, $\rho\sim\omega^\alpha$, then the
exponent $z$ depends on $\alpha$.
For instance, the pure system of Dirac fermions has $\alpha=1$ and
there is an effective exponent $\alpha=1/3$ in the case of a random mass,
according to the discussion below Eq. (\ref{cond3}) (and see also Fig. 3). 
Then Eq. (\ref{D'0}) implies $D\sim \omega^{-\alpha}$, and
we obtain from a simple calculation for the two-particle Green's
function of Eq. (\ref{cricor}), with $k\sim L^{-1}$ and $\eta=0$

\begin{equation}
{\tilde C}(L^{-1},\omega)=\omega^{\alpha-1}{\bar C}(L^{2/(1+\alpha)}\omega),
\end{equation}
where ${\bar C}$ is the scaling function.
With $\alpha=1$ (pure system) this implies $z=1$ and with $\alpha=1/3$ 
it gives $z=3/2$. The above scaling form
does not describe the asymptotic behavior for $\omega\sim0$, but rather the
effective behavior probably relevant for most of the numerical calculations.

The result in Eq. (\ref{cond3}) is also different from the Drude theory 
\cite{ashcroft}, suitably adapted to accomodate the presence of the Lorentz 
force on the scattered electrons:

\begin{equation}
\sigma_{xx}(\omega)=\sigma_0 \frac{1-i\omega\tau}{(1-i\omega\tau)^2
+\omega_c^2\tau^2},
\end{equation}

\noindent
where the zero field DC Drude conductivity $\sigma_0$ and the cyclotron 
frequency $\omega_c$ are defined in the usual way ($\tau$ is a collision time)

\begin{equation}
\sigma_0=\frac{n e^2\tau}{m}, \qquad \omega_c=\frac{eB}{mc}.
\end{equation}

The frequency scale in our study is determined by the value of the hopping
parameter $t$ in the tight-binding model of Ludwig et al. \cite{zie6}. The
physical frequency ${\bar\omega}$ is related to the dimensionless frequency
$\omega$ via $\omega=\hbar{\bar\omega}/t$. Moreover, if we employ
the parameter $b= 2\pi m_c^2\omega/4g$ with the illustrative value of
$m_c=0.01$, the dimensional frequency is related to $b$ via 
${\bar\omega}\approx 4\times 10^{16} b$ Hz for a value of $t=1$ eV.
(Note, however, that our choice $m_c=0.01$ (i.e. $g=0.6$) is probably only
qualitatively significant, since it is difficult to estimate the value of
disorder strength $g$ in a real sample. The above value may be too large
for real systems that have less disorder).  

Our assumption of small frequency breaks down if $\hbar{\bar\omega}$
becomes of
the order of the characteristic energy scale given by the hopping rate $t$.
The latter is of the order of 1 eV in realistic systems; thus, we expect 
a cross-over frequency of about ${\bar\omega}\approx 10^{15}$ Hz. Typical
experiments were performed for frequencies between 0.2 and 14 GHz \cite{engel};
this is well below the cross-over frequency, and our small frequency
approximation should hold.

In a recent paper Shahar et al. \cite{sha2} have studied the behavior
of the longitudinal resistivity, in the neighborhood of the
transition from the QH liquid to the Hall insulator, known to be of the form

\begin{equation}
\rho_{xx}=e^{-\Delta \nu/\nu_0(T)}.
\end{equation}

\noindent
These Authors found a deviation from the conventional scaling form
$\nu_0(T)=T^{1/z\nu}$. Instead, they suggest to fit their experimental data
with the form 

\begin{equation}
\nu_0(T)=\beta+\alpha T,
\end{equation}

\noindent
where $\alpha, \beta>0$ (both parameters $\alpha$ and $\beta$ depending
strongly on sample properties). A very similar result was found by Balaban
et al. \cite{balaban} for the frequency-dependent conductivity at zero
temperature.
These results are consistent with our finding of a deviation of the broadening
from a power law for very low frequencies. It is a consequence of the 
non-zero band width of delocalized states in the disorder-dependent interval
$[-2\lambda\exp(-\pi/g),2\lambda\exp(-\pi/g)]$ present in our model.

In conclusion, we have developed a theoretical treatment for the dynamical
transport properties of the IQH system near the plateau-to-plateau transition.
Our results indicate an increase of the longitudinal conductivity accompanied
by the broadening of the conductivity peak as the frequency is increased.
This behavior follows, at large frequencies, a power law also found in recent
experiments \cite{engel} and numerical studies \cite{avishai}. However, for
very low frequency we found deviations from
a power law, similarly to what was found in recent experiments by Shahar et 
al. \cite{sha2} and Balaban et al. \cite{balaban}. For larger values of the
frequency, the scaling behavior of
the conductivity is recovered with an exponent 1/3 which is in agreement
with the recent work by Avishai and Luck \cite{avishai}. These results 
follow from a theoretical treatment in which the delocalized states responsible
for the IQH transition are properly accounted for.

\begin{center} ACKNOWLEDGEMENTS \end{center}

This work was supported in part (GJ) by EC contract No. ERB4001GT957255.

\begin{center}  FIGURE CAPTIONS \end{center}

Figure 1. Real (upper curve) and imaginary (lower curve) parts of the 
longitudinal conductivity 
$\sigma_{xx}(\omega)$ in units of $e^2/h$ as a function of the dimensionless 
frequency parameter $b=2\pi m_c^2\omega/4g$ (see text).

Figure 2. Broadening of the conductivity peak with frequency. $Re(\sigma_{xx})$
is plotted vs. the Dirac mass $m$ (in units of the hopping parameter $t$)
and the dimensionless frequency parameter $b$.

Figure 3. Average density of states for $m_c=0.01$.

\begin{figure}
\begin{center}\mbox{\psbox{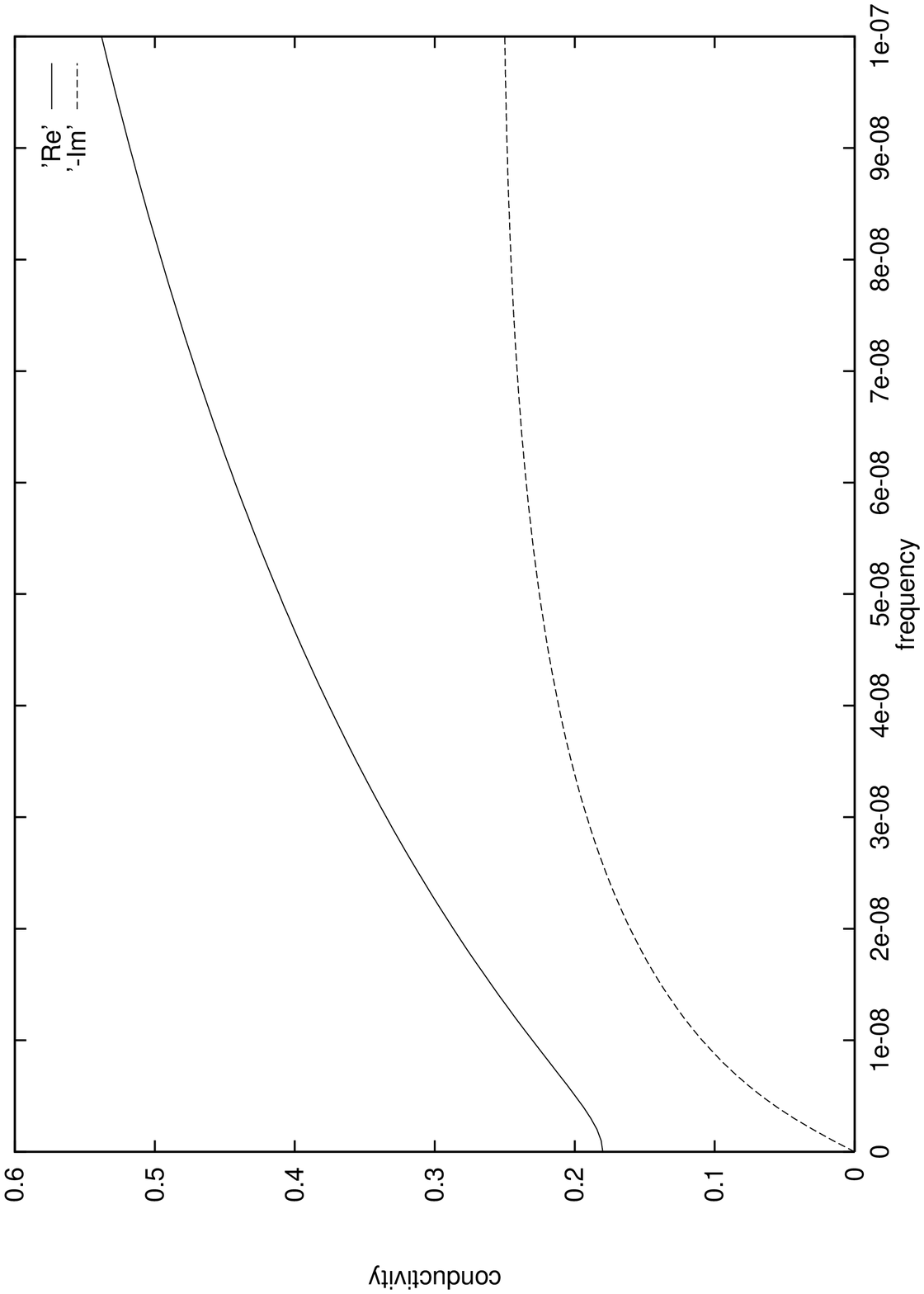}}\end{center}
\end{figure}

\begin{figure}
\begin{center}\mbox{\psbox{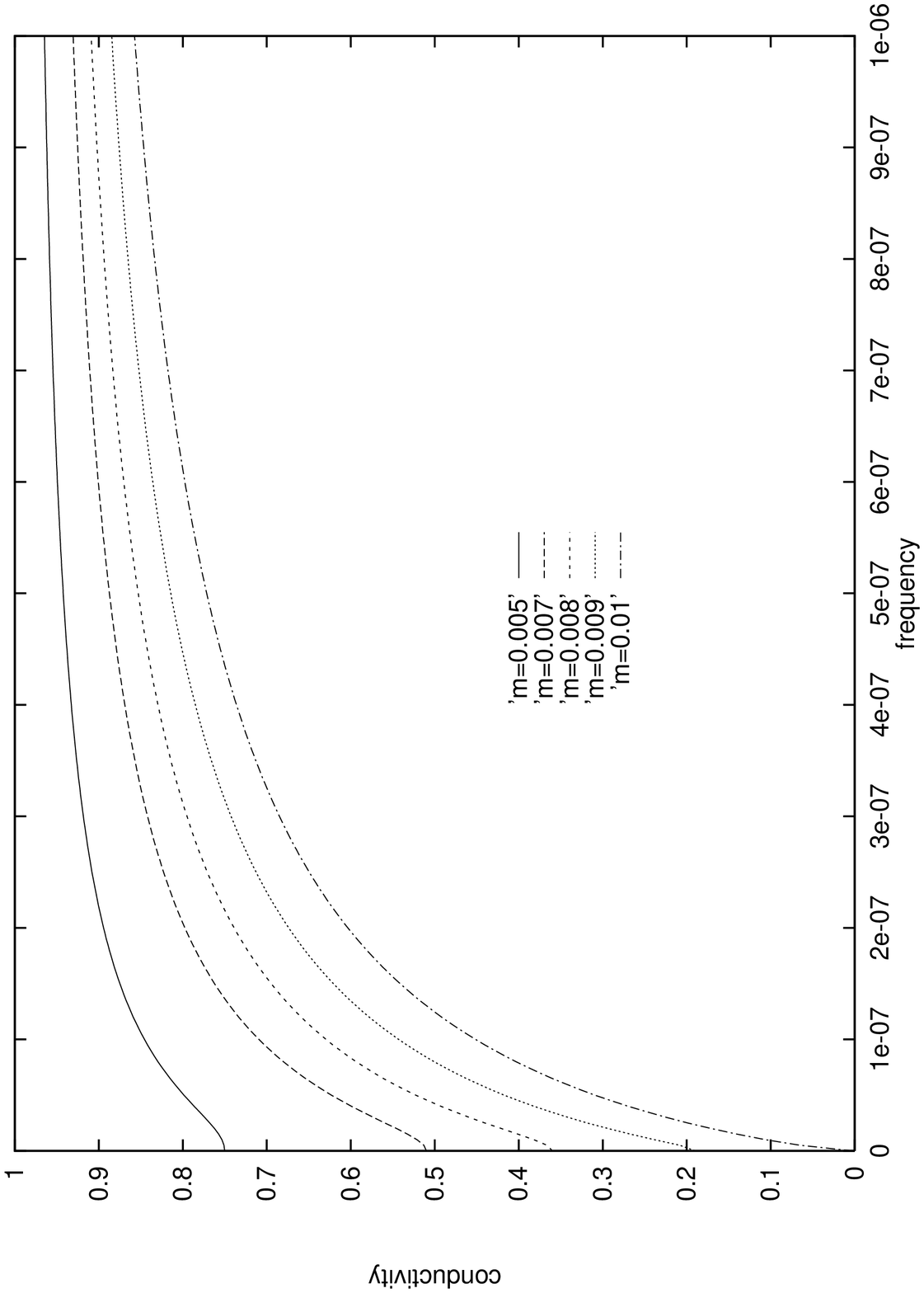}}\end{center}
\end{figure}

\begin{figure}
\begin{center}\mbox{\psbox{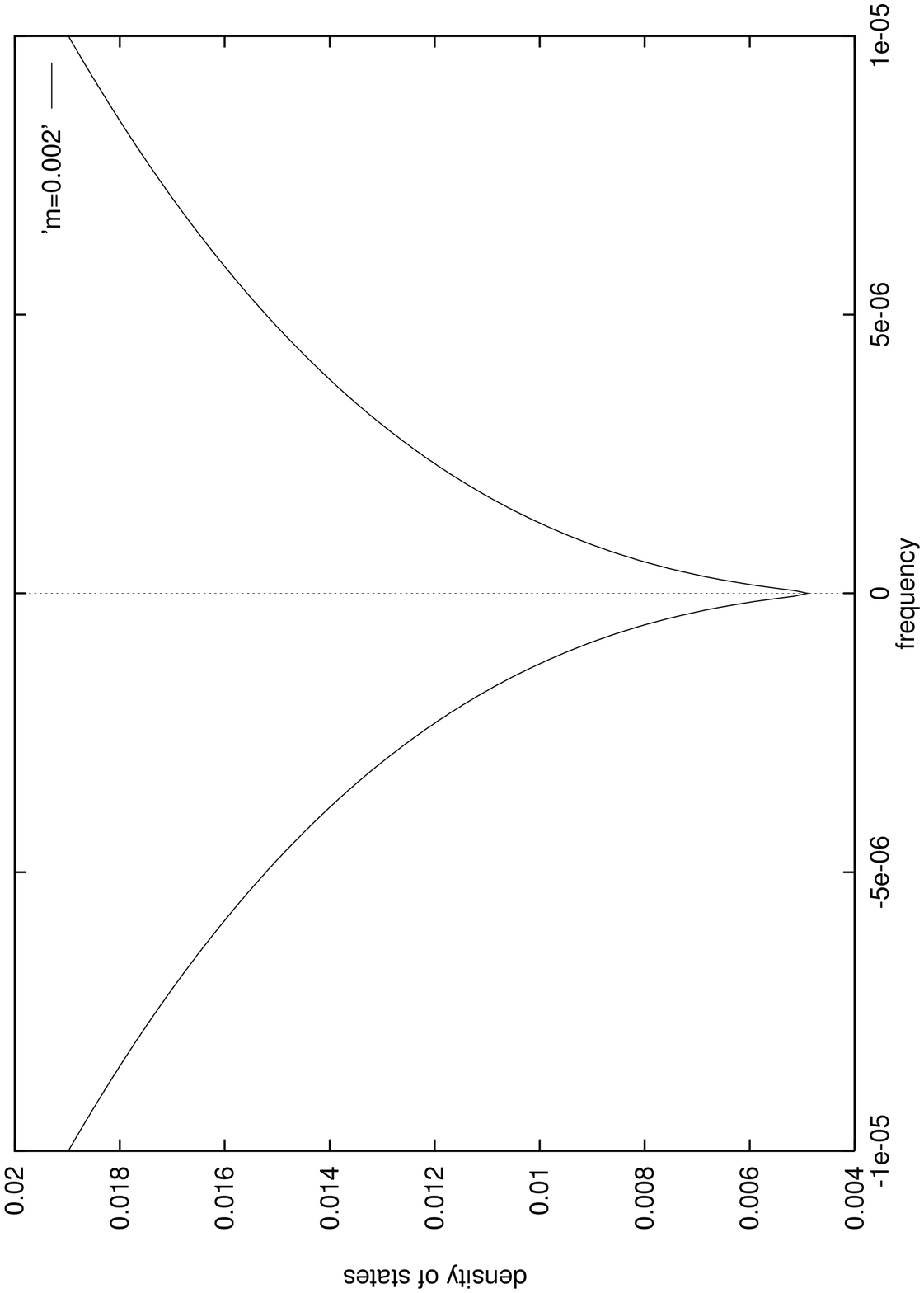}}\end{center}
\end{figure}


\begin{references}
\bibitem{remark}
Random mass and random energy seem to be equally important, in the sense
that properties are qualitatively the same if both or only one of them are
present.
\end{references}
\end{document}